\documentstyle[11pt,aaspp4,flushrt]{article}
\lefthead{Lee \& Park}
\righthead{Symbiotic Star RR Tel}
\begin{document}
\title{Toward the Evidence of the Accretion Disk Emission in the Symbiotic Star 
RR~Tel }
\author{Hee-Won Lee and Myeong-Gu Park}
\affil{Dept. of Astronomy and Atmospheric Sciences, 
Kyungpook National University, Taegu, Korea}
\authoremail{hwlee@vega.kyungpook.ac.kr mgp@kyungpook.ac.kr}
\begin{abstract}
In this paper, we argue that in the symbiotic star RR Tel 
the existence of an accretion disk around the hot companion is strongly
implied by the characteristic features exhibited by the Raman-scattered 
O VI lines around 6830 \AA\  and 7088 \AA. High degrees of polarization and 
double-peaked profiles in the Raman-scattered lines and single-peak profiles 
for other emission lines are interpreted as line-of-sight effects, where the 
H~I scatterers near the giant see an incident double-peaked profile and an 
observer with a low inclination sees single-peak profiles. 
It is predicted that different mass concentrations around the accretion 
disk formed by a dusty wind may lead to the disparate ratios of the blue 
peak strength to the red counterpart observed in the 6830 and 7088 features.
We discuss the evolutionary links between symbiotic stars and bipolar 
protoplanetary nebulae and conclude that the Raman scattering processes
may play an important role in investigation of the physical properties
of these objects.

\end{abstract}
\keywords{stars : symbiotic --- accretion disk --- polarization --- 
scattering --- planetary nebulae --- evolution}

\section{Introduction}

Symbiotic stars exhibit the thermal components typical of a cool star
and a hot star in their spectra with additional emission nebulosity.
It is usually thought that they form a binary system of a giant suffering 
a heavy mass loss and a white dwarf surrounded by an emission nebula
\markcite{ken86}(e.g. Kenyon 1986). 
The double-peaked profiles often observed in the emission lines of many 
symbiotic systems convincingly imply that the emission regions may be 
characterized by disk-type motions. The variability and nova-like outbursts 
may also be attributed to the existence of an accretion disk. Since the giant 
provides material to the hot star in the form of a stellar wind, it is 
plausible to expect that an accretion disk may be formed around the hot star 
\markcite{rob94} (Robinson et al. 1994). Recently, 
\markcite{mm98}Mastrodemos \& Morris (1998) presented their numerical 
computations on disk formation in a wide binary system of a white dwarf 
and a giant through a dusty wind.

An important and distinct aspect of spectroscopy of symbiotic stars is 
provided by the Raman-scattered features around 6830 \AA\ and 7088 \AA, 
which are identified by \markcite{hms89} Schmid (1989). According to him, 
they are originally the O VI 1032, 1038 
doublet lines that are absorbed and re-emitted by atomic hydrogen initially in 
the ground $1s$ state and finally to de-excite to the $2s$ state. Basic 
atomic physics of the Raman-scattering processes is discussed by 
many researchers \markcite{ll97a, nsv89, sm69, sd92} (Lee \& Lee 1997a, 
Nussbaumer et al. 1989, Saslow \& Mills 1969, Sadeghpour \& Dalgarno 1992).

The Raman-scattered features are characterized by the strong polarization and 
the Doppler enhancement by a factor of $\sim 7$ attributed to the incoherence 
of scattering. Therefore, the broadened profiles and polarization are 
expected to carry much information about the physical properties of 
the emission regions and scattering geometry. This point is illustrated 
by the numerical works by \markcite{hh97}Harries \& Howarth 
(1997) and \markcite{hms92} Schmid (1992) and by the spectropolarimetric 
observations performed by \markcite{hh96, ss90} Harries \& Howarth (1996), 
and Schmid \& Schild (1990).

However, thus far it appears that neither numerical works nor observational 
reports successfully address the important role of the Raman-scattered features
that they provide a unique view of the emission region from the neutral 
scatterers rather than in the line of sight to the observer. One good 
example of this point is found in the spectropolarimetry of the broad 
double-peaked H$\alpha$ 
line in the radio galaxy Arp 102B by \markcite{cor98} Corbett et al. (1998).
They proposed that the single-peaked profile in the polarized flux is mainly 
associated with the different line of sight to the scatterers than the 
observer's sight line that is prone to give a double-peaked profile in the 
direct flux. Similar but more interesting effects are expected 
in the case of the Raman-scattered lines, because these are purely scattered 
features with no direct component and furthermore the associated polarized 
fluxes carry additional and complementary information.

Recently, simultaneous observations of the Raman-scattered features and 
emission lines in the UV range have been performed \markcite{esp95, bes98, 
hms98}(Espey et al. 1995, Birriel et al. 1998, Schmid 1998). In this regard, 
this viewpoint about the Raman-scattered lines is expected to be 
useful for building more elaborate models of symbiotic stars.
In this Letter, we adopt this point to compute the profiles and polarization
of the Raman-scattered O VI lines and explain many prominent features
seen in the symbiotic star RR~Tel, and discuss the evolutionary status in 
relation to bipolar protoplanetary nebulae. 

\section{Model}

\subsection{Accretion Disk in Symbiotic Stars}
In this subsection, we give a brief description of our model adopted for the 
computation of the profiles and polarization of the Raman-scattered lines
in the symbiotic star RR Tel. Spectropolarimetric observations of RR~Tel 
have been performed by
a number of researchers \markcite{hh96, esp95} (e.g. Harries \& Howarth 1996,
Espey et al. 1995). The main features of the spectra of RR~Tel include
the clear double-peaked profiles, high degree of polarization
and polarization flip in the red wing in the Raman-scattered lines.
Because the scatterers responsible for the Raman-scattered lines
are believed to be located near the giant, the double-peaked profiles
imply that the emission regions are characterized by a disk-type motion.

\begin{figure}
\plotone{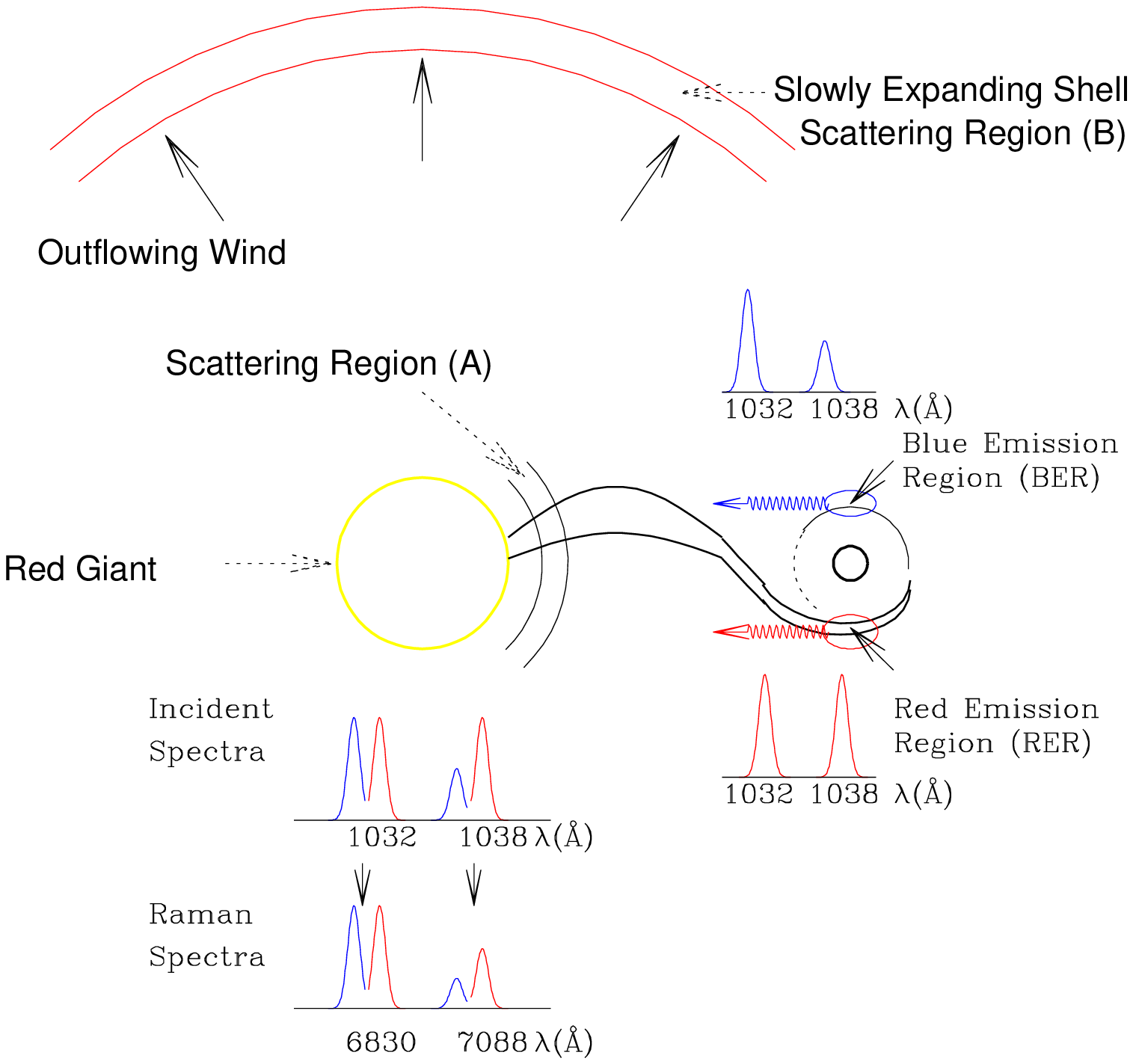}
\caption{A schematic diagram illustrating the accretion disk emission
region and the scattering geometry for the symbiotic star RR Tel
adopted in this paper. Conceptually there are two emission 
regions around the hot star denoted by `RER' and `BER', which provides 
red-shifted line photons and blue-shifted ones to the direction of the giant
respectively.  
The scattering region is largely divided into two regions, i.e., region (A) 
near the giant that is very thick  and region (B) that is a large spherical 
shell expanding slowly with speed $v_{shell} = 30\ {\rm km\ s^{-1}}$ and 
total scattering optical depth $\tau_T = 0.3$. The scatterers near the giant 
see a double peak incident source whereas the source appears to have a 
single-peaked profile.  It is assumed that when the incident wavevector 
${{\bf\hat k}_i}$ makes an angle less than $45^\circ$ with the binary axis, 
then the photon hits region (A) and otherwise it enters region (B).
\label{fig1}}
\end{figure}

\placefigure{fig1}
In Fig. 1 is shown a schematic diagram illustrating the accretion disk
emission regions and the scattering geometry adopted in this Letter. There
are two main emission regions around the hot star, that is, the red emission
region (marked by `RER') and the blue emission region (marked by `BER'), which
give the red and blue components in the direction of the binary axis, 
respectively. Here, by the binary axis we mean the line connecting the two
stars. According to \markcite{mm98}Mastrodemos \& Morris (1998) more 
mass concentration is expected in the RER than in the BER. We discuss this 
effect in the next section.

The scatterers are assumed to be concentrated mainly in two regions,
that is, in region (A) near the giant star and in region (B) forming 
a slowly expanding shell. \markcite{sol84}Solf (1984) investigated
the bipolar mass outflow of the symbiotic star HM Sge and found that
in addition to the outflow of velocity $\sim 200 \ {\rm km\ s^{-1}}$,
there are slowly moving features near the equatorial plane. The amount
of neutral hydrogen in region (B) is not certain at the moment, and in 
this paper it is assumed to be much smaller than that in region (A), but
not negligible. It is also assumed that when the incident wavevector 
${{\bf\hat k}_i}$ makes an angle less than $45^\circ$ with the binary axis, 
then the photon hits region (A) and otherwise it is scattered in region (B).

\subsection{Double-Peaked Emission from an Accretion Disk}

\begin{figure}
\plotone{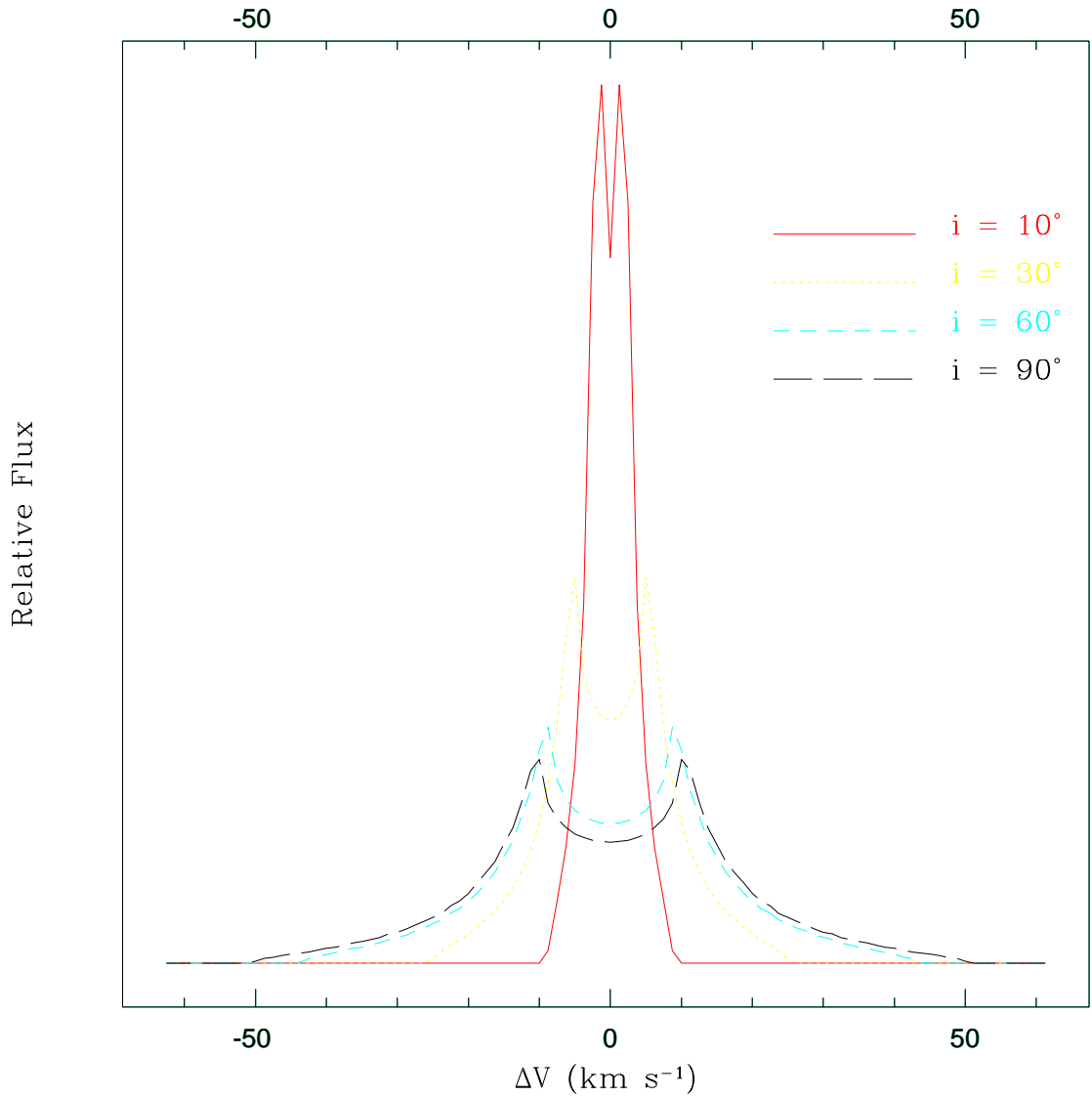}
\caption{The emission line profiles from an optically thick accretion disk 
for inclination angles $i= 10^\circ$(solid line), 30$^\circ$(dotted line), 
60$^\circ$(dashed line), and 90$^\circ$(long-dashed line)
obtained by using the approximation adopted by Horne \& Marsh (1986). Note
that $i=90^\circ$ corresponds to the edge-on disk case.
The maximum velocity of the emission region is chosen to be 
$50\ {\rm km\ s^{-1}}$ and the thermal broadening has not been applied.
\label{fig2}}
\end{figure}

\placefigure{fig2}
We follow the approximation adopted  by \markcite{hm86}Horne \& Marsh (1986) 
to generate the line profiles from the disk inclined at various angles. The 
emission lines are assumed to be optically thick, and the disk obeys the 
Keplerian rotation. We also assume that the disk is geometrically thin with 
the disk height ratio $H/R = 1/50$. Since we do not know enough about the 
accretion disk in symbiotic stars and even less about the emission regions 
in the disk, we just take the line source function to be a simple
power-law $S_L \propto r^{-1.2}$, which has been used by \markcite{hm86}Horne 
\& Marsh (1986) to match the line shape of H$\alpha$ in Z Cha. 
Fig. 2 shows the line profiles for inclination angles $i= 10^\circ$, 
30$^\circ$, 60$^\circ$, 90$^\circ$. The maximum velocity of the emission 
region is chosen to be $50\ {\rm km\ s^{-1}}$ and the thermal broadening 
has not been applied. We note that the low-inclination profile is narrower 
than the high-inclination counterpart.

In this paper, we do not attempt 
to reproduce the observational results by fitting accurately, and instead 
we want to point out the physical origins of the main characteristic features 
shown in the observational data. Therefore, for the computation of the 
generic profiles and polarization, we use two Gaussians to approximate the 
profiles in Fig.~2, and adopt the peak strength ratio as a new free parameter.
It is referred to \markcite{ll97b}Lee \& Lee (1997b) for the detailed 
description of the Monte Carlo computation of the scattered flux profile and 
polarization. (See also \markcite{hh97, hms92}Harries \& Howarth 1997, 
Schmid 1992) 

\section{Result and Discussion} 

\subsection{Profile and Polarization}

\begin{figure}
\plotone{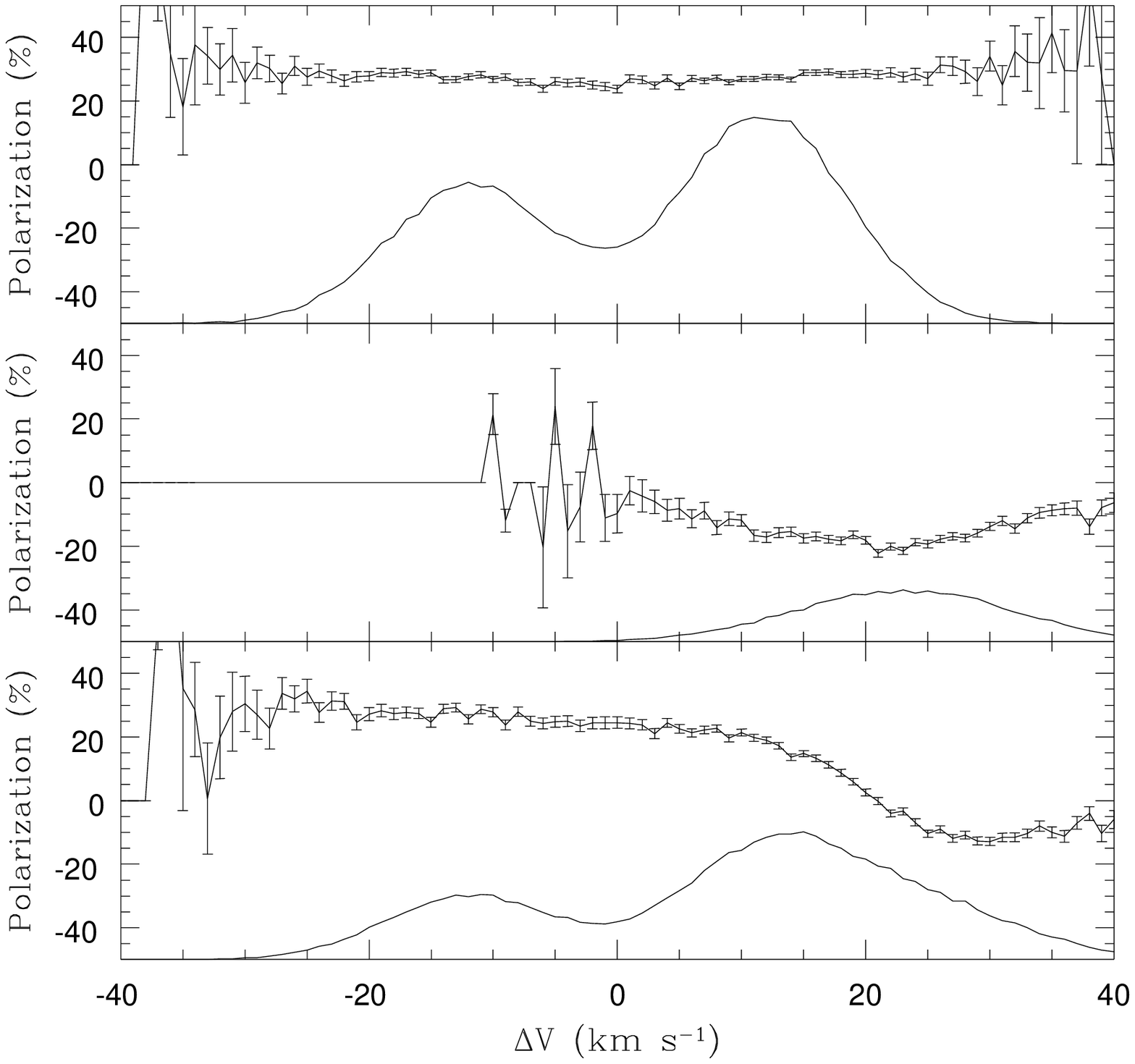}
\caption{ Profiles (dotted lines) and polarization (solid lines with 1$\sigma$
error bars) of the Raman-scattered O VI lines
from accretion disk type emission regions shown in Fig. 1. 
In Panel (a), the flux and polarization of the component scattered 
in region (A) of Fig. 1 are shown.
In Panel (b), the same quantities as in Panel a for the component scattered
in region (B) are shown.
In Panel (c) are shown the synthetic profile and polarization of the quantities
shown in Panels (a) and (b).
\label{fig3}}
\end{figure}

\placefigure{fig3}
In Fig. 3 is shown the main result. We set the strength ratio of the blue
part to that of the red part to be 2/3 and discuss this point in the next 
subsection. The top panel shows the
flux and polarization of the component scattered near the giant. 
In Panel (b) the same quantities are shown for the radiation scattered
in the spherical shell receding with velocity $v_{shell} = 30 \ {\rm km\ 
s^{-1}}$ and total scattering optical depth $\tau_{T} = 0.3$. In the bottom
panel is shown the sum of the preceding two components. Here, the 
polarization direction is represented by the sign, where the positive
sign represents the polarization in the direction perpendicular to the
binary axis whereas the negative sign the parallel direction.

The double-peaked profile in the top panel is obtained because the
scatterers near the giant see a double-peaked incident source. This
component is strongly polarized in the direction perpendicular to
the binary axis. Because the relative velocity of the scatterers near
the giant is ignored, the scattered flux profile is almost the same
as the incident profile in the binary axis direction and the breadth
of the feature represents mainly the kinematics of the source part.

In Panel (b), we obtain
a broad profile with a single peak. The polarization is in the direction
parallel to the binary axis and is weaker because region (B) subtends a
larger solid angle than region (A) does. Due to the relative motion of the 
shell, the location of the peak is redshifted from the center. The single-peaked
profile is obtained because the averaged profile incident to the shell
is smooth with a single peak near the center.

In combining the two components we may see interesting fine points that
may be revealed in the spectropolarimetry of symbiotic stars.
Firstly, the strength of the scattered component is determined mainly
by the scattering optical depth and the solid angle of the scattering
region. Our choice is such that the shell has a small scattering optical
depth so that the synthetic scattered flux is dominantly determined
by the near axis-scattered component. In RR Tel, most emission lines
are single-peaked and the Raman lines show the double-peaked profile,
which strongly imply the emission regions in disk-type motion
with low inclination.

Secondly, the shell-scattered component is red-shifted
and single-peaked. Therefore, this component adds more flux to the red part
and increases slightly the ratio of the red peak strength to that of the 
blue in the synthetic flux. However, since the shell-scattered component
is polarized in the parallel direction, in the overlapping region with
the flux scattered in region (A) the total polarized flux is reduced as
a result of cancellation of polarization but remains still in the 
perpendicular direction.
Because the shell-scattered component extends more redward than the 
component scattered in region (A), 
there remains a parallelly polarized flux in the reddest part where 
the (A)-scattered component does not contribute. The polarization
flip in the red wing part is an important feature shown in many 
spectropolarimetric observations of symbiotic stars.

Finally, as \markcite{hh96}Harries \& Howarth (1996) pointed out, the overall
profiles observed for many symbiotic systems are broader than the terminal 
speed of the stellar wind associated with K or M type giants and the commonest 
profile type is triply-peaked where the reddest component is often polarized 
oppositely compared with the remainder part. It turns out that the overall 
breadth of the profile is determined by the kinematics of the source part and 
the relative motion of the scatterers. Furthermore, by increasing the
speed of the receding shell, we can also generate triply-peaked profiles with 
the polarization flipped in the reddest component. 

\subsection{O VI Doublet Strength Ratio}

In this subsection we discuss a possible prediction of the symbiotic
star model consisting of an accretion disk with a bipolar wind.
According to \markcite{mm98}Mastrodemos \& Morris (1998), it is expected 
that the RER has a higher optical depth than the BER does. For the O VI 1034
doublet, the 1032 photons have twice larger optical depth than the 1038 
counterparts. In an optically
thin medium, the oscillator strength ratio becomes that of the emergent
line strength ratio. However, in an optically thick medium the emergent 
line strength of the 1032 \AA\ component becomes similar to that of the 
1038 \AA\ component. However, complications may occur in
a non-stationary medium, where the sensitive dependence of the resonance
scattering cross section on the velocity field may easily alter the
escape probability so that it becomes eventually proportional to the velocity
gradient \markcite{sob63, lb97, mic 88}(e.g. Sobolev 1963, Lee \& 
Blandford 1997, Michalitsianos et al. 1988). Further complications
can be expected if the medium is dusty, in which case resonantly scattered
O VI photons are subject to destructions by the dust particles
\markcite{neu91}(e.g. Neufeld 1991).

It is an interesting possibility that if the BER is optically thin
and the RER is thick, then the Raman-scattered line around 6830 \AA\
will show a larger ratio of the blue peak strength to the red peak strength
than the 7088 \AA\ feature, as is depicted in Fig. 1. It appears that in RR Tel
the 7088 \AA\ feature has a weaker red part compared that of the 6830 \AA\
feature (Harries \& Howarth 1997). However, since the 7088 \AA\ feature is
much weaker, higher resolution spectroscopy with good signal to noise ratio 
must be invoked to exclude other possibilities such as selective interstellar 
absorptions.

\markcite{hn86}Hayes \& Nussbaumer (1986) proposed that 
the electron density $n_e \sim 3\times 10^6 \ {\rm cm^{-3}}$ and the size
of the emission line region $R\sim 10^{15}\ {\rm cm}$ in RR Tel.
A simple computation of the line center optical depth of the O VI 1034 
doublet in the emission region gives
$$ \tau_c \sim 7.2\times 10^{4} T_4^{-1/2} [f_i A_{O VI}/10^{-4}]
[n_e/(10^6 \ {\rm cm^{-3}})] [R/(100\ {\rm AU})], \eqno(3.1)$$
where $T_4$ is the temperature in units of 
$10^4\ {\rm K}$, $A_{O VI}$ is the O VI fraction in number and $f_i$ is
the oscillator strength of the O VI resonance transition \markcite{rl79}
(Rybicki \& Lightman 1979).
Apparently, the emission region in RR Tel is very optically thick and
the O VI doublet strength ratio would be nearly 1:1 in the entire region, 
even if the BER is an order of magnitude rarer in mass concentration
than the RER. Therefore, both high resolution spectroscopy and further
theoretical work on the radiative transfer based on a more refined model
will shed light on this point.

\subsection{Evolutionary Status of Symbiotic Stars}

Asymmetric morphologies in the emission nebulae have been known in many 
symbiotic systems including V 1016 Cyg \markcite{ss96}(Schild \& Schmid 1996), 
HM Sge \markcite{sol83, eyr95}(Solf 1983, Eyres et al. 1995).
\markcite{sol84}Solf(1984) emphasized that a large fraction of symbiotic 
stars exhibit bipolarity and that there is a remarkable morphological 
resemblance with postnova shells and protoplanetary nebulae.
The spectroscopic similarity between symbiotic stars and bipolar 
protoplanetary nebulae has led many researchers to propose evolutionary 
links between them \markcite{it96, cor95, cs95}(e.g. Iben \& Tutukov 1996, 
Corradi 1995, Corradi \& Schwarz 1995). A supporting argument to this effect 
is that the binarity 
may play an important role in forming an aspherical morphology of the nebula, 
typically characterized by bipolarity \markcite{mor87,sok98}(Morris 1987,
Soker 1998). However, the binarity of symbiotic stars is well-established 
whereas that of the bipolar protoplanetary nebulae still remains controversial.

The accretion disk formation in a young planetary nebula has been discussed
by a number of researchers \markcite{mor87, sl94}(e.g. Morris 1987, 
Soker \& Livio 1994). Recently, \markcite{mm98}Mastrodemos \& Morris (1998) 
presented an interesting numerical computation on the dusty wind 
accretion in a detached binary of a mass-losing AGB star and a hot star 
that may be responsible for the bipolar morphology in protoplanetary nebulae. 
They found that a permanent and stable accretion disk is formed around 
the hot companion with an efficient cooling associated with dust in the wind
(see also \markcite{tj93} Theuns \& Jorissen 1993). In particular, they
concluded tha the limiting binary separation for disk formation should be
greater than 20 AU for their M4 model. Therefore, it is interesting
that the evolutionary links between symbiotic stars and bipolar protoplanetary 
nebulae imply that the hypothesis of accretion disk emission in RR Tel
is also rendered a strong case. Furthermore, as shown in Fig.~1 the
bipolar wind along the disk axis may provide natural scattering sites for the
Raman-scattered flux that constitutes the red wing part with polarization flip.

The Raman-scattered lines are characterized by the high polarization and 
broadened profiles, which enable one to put strong constraints on
the scattering geometry. Therefore, the main spectroscopic and polarimetric
features imply very plausibly that the symbiotic star RR Tel may possess
an accretion disk with a bipolar outflow. So called `type 2' symbiotic
stars seem to show similar behaviors in the Raman scattered lines
and exhibit bipolar type morphologies.  Furthermore, about a half of 
symbiotic stars exhibit the Raman scattered lines. 
Fine-tuned conditions such as a restricted range of the initial mass ratios 
of the two constituent stars and the orbital parameters affecting the mass 
transfer rate may be needed for the co-existence of a large amount of neutral 
hydrogen and highly ionized nebulae which characterize the symbiotic 
phenomenon.

An interesting example is provided by \markcite{peq97} P\'equignot (1997), 
who performed high resolution spectroscopy on a young planetary nebula 
NGC~7027 and found a Raman scattered He II line blueward of H$\beta$.
Gurzadyan (1996) discusses the relation of NGC~7027 to symbiotic stars,
noting that NGC~7027 shows high ionization lines along with a strong IR
component. So far the Raman-scattering by H I is found to operate only 
in symbiotic stars except NGC~7027, and theoretical possibilities have been 
discussed in other astrophysical objects such as active galactic nuclei
\markcite{nsv89, ly98}(Nussbaumer et al. 1989, Lee \& Yun 1998). 
Therefore, it will remain an interesting possibility that similar processes
are expected to operate in bipolar protoplanetary nebulae, in which case
the Raman scattering can be regarded as an important tool that reveals
the evolutionary links between symbiotic systems and bipolar protoplanetary
nebulae.

\acknowledgements
We are grateful to  Dr. Hwankyung Sung and Sang-Hyeon Ahn for helpful 
discussions. We also thank the referee, who gave many suggestions that improve
the presentation of this paper. HWL is supported by the Post-Doc. program 
(1998) at the Kyungpook National University. MGP gratefully acknowledges 
support from KOSEF grant 971-0203-013-2.

\clearpage
 
\begin{references}

\reference{bes98} Birriel, J. J., Espey, B. R. \& Schulte-Ladbeck, R. E.,
1998, ApJ, 507, L75

\reference{cor98} Corbett, E., Robinson, A., Axon, D., Young, S., Hough, J. 
1998, MNRAS, 296, 721

\reference{cor95} Corradi, R. L. M., 1995, MNRAS, 276, 521

\reference{cs95} Corradi, R. L. M., \& Schwarz, H. E. 1995, A\&A, 293, 871.

\reference{esp95} Espey, B. R., Schulte-Ladbeck, R. E., Kriss, G. A.,
Hamann, F., Schmid, H. M., Johnson, J. J., 1995, ApJ, 454, L61

\reference{eyr95} Eyres, S. P. S., Kenny, H. T., Cohen, R. J., Lloyd, H. M.,
Dougherty, S. M., Davies, R. J., Bode, M. F., 1995, MNRAS, 274, 317

\reference{gur96} Gurzadyan, G. A., 1996, The Physics and Dynamics of 
Planetary Nebulae, Springer-Verlag, Berlin

\reference{hh96} Harries, T. J., \& Howarth, I. D., 
1996, A \& AS, 119, 61

\reference{hh97} Harries, T. J., \& Howarth, I. D., 
1997, A \& AS, 121, 15

\reference{hn86} Hayes, M. A. \& Nussbaumer, H., 1986, A\& A, 161, 287

\reference{hm86} Horne, K. \& Marsh, T. R., 1986, MNRAS, 218, 761

\reference{it96} Iben, JR., I. \& Tutukov, A. V., 1996, ApJS, 105, 145

\reference{ken86} Kenyon, S. J., 1986, The Symbiotic Stars, CUP, Cambridge

\reference{lb97} Lee, H. -W. \& Blandford, R. D., 1997, MNRAS, 288, 19

\reference{ll97a} Lee, H. -W. \& Lee, K. W., 1997, MNRAS, 287, 211

\reference{ll97b} Lee, K. W. \& Lee, H. -W., 1997, MNRAS, 292, 573

\reference{ly98} Lee, H. -W. \& Yun, J. -H, 1998, MNRAS, 301, 193

\reference{mm98} Mastrodemos, N. \& Morris, M., 1998, ApJ, 497, 303

\reference{mic88} Michalitsianos, A. G., Kafatos, M., Fathey, R. P.,
Viotti, R., Cassatella, A., \& Altamore, A., 1988, ApJ, 331, 477

\reference{mor87} Morris, M., 1987, PASP, 99, 1115

\reference{neu91} Neufeld, D. A., 1991, ApJ, 370, L85

\reference{nsv89} Nussbaumer, H., Schmid, H. M.\& Vogel, M.,
1989, A\& A, 211, L27 

\reference{peq97} Pequignot, D., Baluteau, J. -P., Morisset, C., Boisson, C.,
1997, A\& A, 323, 217

\reference{rob94} Robinson, K., Bode, M.F., Skopal, A., Ivison, R. J., \& 
Meaburn, J., 1994, MNRAS 269, 1

\reference{rl79} Rybicki, G.B. \& Lightman, A.P.,1979, Radiative Processes in 
Astrophysics (Wiley-Interscience),

\reference{sm69} Saslow, W. M.,\&  Mills, D. L., 
1969, Physical Review, 187, 1025 

\reference{sd92} Sadeghpour, H. R. \& Dalgarno, A. 1992, J. Phys. B: At. Mol. 
Opt.Phys, 25, 4801 

\reference{ss96} Schild, H. \& Schmid, H. M., 1996, A\& A, 310, 211

\reference{hms89} Schmid, H. M. 1989, A\& A, 211, L31 

\reference{hms92} Schmid, H. M. 1992, A\& A, 254, 224 

\reference{hms98} Schmid, H. M. 1998, Rev. of Modern Astronomy, in press

\reference{ss90} Schmid, H. M. \& Schild, H., 1990, A\& A, 236, L13

\reference{sob63} Sobolev, V. V., 1963, A Treatise on Radiative Transfer,
 D. Van Nostrand Company, Inc., Toronto 

\reference{sl94} Soker, N. \& Livio, M. 1994, ApJ, 421, 219

\reference{sok98} Soker, N. 1998, ApJ, 496, 833

\reference{sol83} Solf, J., 1983, ApJ, 266, L113

\reference{sol84} Solf, J., 1984, A\& A, 139, 296

\reference{tj93} Theuns, T., \& Jorissen, A., 1993, MNRAS, 265, 946

\end{references}
\end{document}